\documentclass[acmsmall,nonacm]{acmart}
\title[Assessing Routing Algorithms for PCNs: Full Evaluation Results]{Assessing Routing Algorithms for PCNs: Full Evaluation Results}
\author{David Lobmaier}
\affiliation{\institution{TU Wien}
	\city{Vienna}
	\country{Austria}}
\author{Rafael Konlechner}
\affiliation{\institution{TU Wien}
	\city{Vienna}
	\country{Austria}}
\author{Stefan Schulte}
\orcid{0000-0001-6828-9945}
\affiliation{\institution{Christian Doppler Laboratory for Blockchain Technologies for the Internet of Things, Institute for Data Engineering, Hamburg University of Technology}
\city{Hamburg}
\country{Germany}}
\email{stefan.schulte@tuhh.de}
\author{Ingo Weber}
\orcid{0000-XXXX-YYYY-ZZZZ}
\affiliation{\institution{Technical University of Munich \& Fraunhofer Gesellschaft}
	\city{Munich}
	\country{Germany}}
\email{first.last@tum.de}
\setcopyright{none}
\usepackage{url}
\usepackage{array}
\usepackage{natbib}
\usepackage{subcaption}
\usepackage{multirow}   
\usepackage{graphicx}
\usepackage[disable]{todonotes}
\usepackage{longtable}
\usepackage[ruled,linesnumbered]{algorithm2e}

\SetKw{Break}{break}

\newcolumntype{H}{>{\setbox0=\hbox\bgroup}c<{\egroup}@{}}

\begin{document}
\begin{abstract}
Within this Technical Report, we present the full analysis of 61 routing protocols for Wireless Sensor Networks (WSNs) for the purposes of routing in Payment Channel Networks (PCNs). In addition, we present the full results of the implementation of the three algorithms E-TORA, TERP, and M-DART.
\end{abstract}

\maketitle

\section{Candidate Algorithms} \label{design:routing:candidates}

Routing algorithms in WSNs have been a popular topic in recent years.
As a result, there are many approaches available in the literature.
In this supplementary material to the paper ``Assessing Routing Algorithms for Payment Channel Networks'', we classify a collection of WSN algorithms, based on their ability to scale and self-organize, trustlessness, and communication compatibility.
Then, we filter them for the most promising candidates in the context of PCN routing.

Our collection of WSN routing algorithms is mainly based on the work of Pantazis et al.~\cite{Pantazis2013EnergyERP} that has been extended with more recent approaches.
Also, a missing algorithm category that uses DHTs has been added.
The complete list can be seen in Table~\ref{design:routing:wsn-protocols}, where each algorithm is assessed to fulfill ($\bullet$), partly fulfill ($\circ$) or fail to meet ($\times$) the abovementioned requirements.

With regard to the utilization of WSN routing algorithms in the field of PCNs, we begin by excluding algorithms that do not meet the requirement for scalability and resource efficiency.
By using routing tables for all nodes of the network or flooding WRP~\cite{01WRP}, TBRPF~\cite{02TBRPF}, HPAR~\cite{18HPAR}, MWE~\cite{28SWEMWE}, DREAM~\cite{34DREAM} and LMR~\cite{47LMR} can be removed from the list of candidates.
Gossiping, RR~\cite{06RR}, DD~\cite{25DD} and ACQUIRE~\cite{27ACQUIRE} have a risk to easily degenerate into flooding.
Also, the algorithms SWE~\cite{28SWEMWE} and MERR~\cite{40MERR} that need a central control instance are not suited for PCNs.

The next set of routing algorithms are improper for use in PCNs, because of not satisfying the self-organization and flexibility requirement.
VGA~\cite{14VGA}, SPastry~\cite{61SPastry} and VCP~\cite{62VCP} have a fixed topology after initialization.
For cluster forming, the algorithms BCDCP~\cite{16BCDCP}, SHPER~\cite{23SHPER}, DHAC~\cite{24DHAC} and COUGAR~\cite{26COUGAR} need a central base station.
Furthermore, non-deterministic routes and not guaranteed delivery of messages disqualifies SPEED~\cite{55SPEED} and its multi-path successor MMSPEED~\cite{56MMSPEED}.
The absence of a global synchronization possibility makes the Sleep/Wake~\cite{19SleepWake} algorithm unusable.

Another requirement that excludes multiple routing algorithms is the communication compatibility.
A prominent member is the LEACH~\cite{09LEACH} protocol and its successors LEACH-C~\cite{10LEACHC}, PEGASIS~\cite{11PEGASIS}, MIMO~\cite{17MIMO}, ELCH~\cite{21ELCH} and SEECH~\cite{59SEECH} that need a shared medium for proper operation.
TEEN~\cite{12TEEN} and APTEEN~\cite{13APTEEN} communicate only if a certain threshold of a sensor value is reached.
A traffic pattern with a central sink is used by TTDD~\cite{15TTDD}, GBDD~\cite{20GBDD}, GRAB~\cite{48GRAB}, HMRP~\cite{49HMRP}, DGR~\cite{51DGR}, DCF~\cite{52DCF}, RPL~\cite{53RPL}, SAR~\cite{54SAR} and MGR~\cite{57MGR}.
Finally, SPIN-PP/SPIN-EC/SPIN-BN/SPIN-RL~\cite{30SPIN} are data-centric algorithms and MIP~\cite{44MIP} and IEMF/IEMA~\cite{45IEMFIEMA} use an agent, i.e., a program that can be migrated between nodes to perform tasks, e.g., value aggregation.

\begin{table}[ht]
	\caption{Classification of WSN Routing Algorithms}
	\label{design:routing:wsn-protocols}
	\tiny
	\centering
	\begin{tabular}{lp{3.2cm}p{3cm}p{1.5cm}p{3cm}H}
		\hline
		&
		\textbf{Scalability \& Resource Efficiency} &
		\textbf{Self-organization \& Flexibility} &
		\textbf{Trustlessness} &
		\textbf{Communication Compatibility} &
		\textbf{Multipath} \\ \hline
		WRP        \cite{01WRP}       & $\circ$   & $\bullet$ & $\circ$   & $\times$  & $\times$  \\ \hline
		TBRPF      \cite{02TBRPF}     & $\times$  & $\bullet$ & $\circ$   & $\bullet$ & $\bullet$ \\ \hline
		TORA       \cite{03TORA}      & $\circ$   & $\circ$   & $\bullet$ & $\bullet$ & $\bullet$ \\ \hline
		Gossiping  \cite{Pantazis2013EnergyERP} & $\circ$   & $\bullet$ & $\circ$   & $\bullet$ & $\times$  \\ \hline
		Flooding   \cite{Pantazis2013EnergyERP} & $\times$  & $\bullet$ & $\circ$   & $\bullet$ & $\bullet$ \\ \hline
		RR         \cite{06RR}        & $\bullet$ & $\circ$   & $\circ$   & $\bullet$ & $\times$  \\ \hline
		E-TORA     \cite{07ETORA}     & $\circ$   & $\circ$   & $\bullet$ & $\bullet$ & $\bullet$ \\ \hline
		ZRP        \cite{08ZRP}       & $\circ$   & $\bullet$ & $\bullet$ & $\bullet$ & $\times$  \\ \hline
		LEACH      \cite{09LEACH}     & $\bullet$ & $\times$  & $\circ$   & $\times$  & $\times$  \\ \hline
		LEACH-C    \cite{10LEACHC}    & $\bullet$ & $\times$  & $\circ$   & $\times$  & $\times$  \\ \hline
		PEGASIS    \cite{11PEGASIS}   & $\bullet$ & $\times$  & $\circ$   & $\times$  & $\times$  \\ \hline
		TEEN       \cite{12TEEN}      & $\bullet$ & $\times$  & $\circ$   & $\times$  & $\times$  \\ \hline
		APTEEN     \cite{13APTEEN}    & $\bullet$ & $\times$  & $\circ$   & $\times$  & $\times$  \\ \hline
		VGA        \cite{14VGA}       & $\bullet$ & $\times$  & $\circ$   & $\times$  & $\times$  \\ \hline
		TTDD       \cite{15TTDD}      & $\circ$   & $\circ$   & $\circ$   & $\times$  & $\bullet$ \\ \hline
		BCDCP      \cite{16BCDCP}     & $\circ$   & $\times$  & $\circ$   & $\bullet$ & $\times$  \\ \hline
		MIMO       \cite{17MIMO}      & $\bullet$ & $\times$  & $\circ$   & $\times$  & $\bullet$ \\ \hline
		HPAR       \cite{18HPAR}      & $\times$  & $\times$  & $\circ$   & $\bullet$ & $\times$  \\ \hline
		Sleep/Wake \cite{19SleepWake} & $\bullet$ & $\times$  & $\circ$   & $\bullet$ & $\bullet$ \\ \hline
		GBDD       \cite{20GBDD}      & $\circ$   & $\circ$   & $\circ$   & $\times$  & $\times$  \\ \hline
		ELCH       \cite{21ELCH}      & $\circ$   & $\circ$   & $\circ$   & $\bullet$ & $\times$  \\ \hline
		SHPER      \cite{23SHPER}     & $\bullet$ & $\times$  & $\circ$   & $\bullet$ & $\times$  \\ \hline
		DHAC       \cite{24DHAC}      & $\bullet$ & $\times$  & $\circ$   & $\bullet$ & $\bullet$ \\ \hline
		DD         \cite{25DD}        & $\circ$   & $\circ$   & $\circ$   & $\bullet$ & $\times$  \\ \hline
		COUGAR     \cite{26COUGAR}    & $\circ$   & $\times$  & $\circ$   & $\times$  & $\times$  \\ \hline
		ACQUIRE    \cite{27ACQUIRE}   & $\circ$   & $\circ$   & $\circ$   & $\bullet$ & $\times$  \\ \hline
		SWE        \cite{28SWEMWE}    & $\times$  & $\times$  & $\times$  & $\times$  & $\times$  \\ \hline
		MWE        \cite{28SWEMWE}    & $\times$  & $\times$  & $\times$  & $\times$  & $\bullet$ \\ \hline
		SPIN-PP    \cite{30SPIN}      & $\bullet$ & $\bullet$ & $\bullet$ & $\times$  & $\times$  \\ \hline
		SPIN-EC    \cite{30SPIN}      & $\bullet$ & $\bullet$ & $\bullet$ & $\times$  & $\times$  \\ \hline
		SPIN-BN    \cite{30SPIN}      & $\bullet$ & $\times$  & $\bullet$ & $\times$  & $\times$  \\ \hline
		SPIN-RL    \cite{30SPIN}      & $\bullet$ & $\bullet$ & $\bullet$ & $\times$  & $\times$  \\ \hline
		DREAM      \cite{34DREAM}     & $\times$  & $\bullet$ & $\circ$   & $\bullet$ & $\times$  \\ \hline
		GEAR       \cite{35GEAR}      & $\circ$   & $\bullet$ & $\circ$   & $\bullet$ & $\times$  \\ \hline
		GEM        \cite{36GEM}       & $\bullet$ & $\bullet$ & $\circ$   & $\bullet$ & $\times$  \\ \hline
		IGF        \cite{37IGF}       & $\circ$   & $\bullet$ & $\circ$   & $\bullet$ & $\times$  \\ \hline
		SELAR      \cite{38SELAR}     & $\circ$   & $\circ$   & $\circ$   & $\bullet$ & $\times$  \\ \hline
		GDSTR      \cite{39GDSTR}     & $\circ$   & $\circ$   & $\circ$   & $\bullet$ & $\times$  \\ \hline
		MERR       \cite{40MERR}      & $\times$  & $\circ$   & $\times$  & $\times$  & $\times$  \\ \hline
		OGF        \cite{41OGF}       & $\bullet$ & $\bullet$ & $\circ$   & $\bullet$ & $\times$  \\ \hline
		PAGER-M    \cite{42PAGERM}    & $\bullet$ & $\bullet$ & $\circ$   & $\bullet$ & $\times$  \\ \hline
		HGR        \cite{43HGR}       & $\bullet$ & $\bullet$ & $\circ$   & $\bullet$ & $\times$  \\ \hline
		MIP        \cite{44MIP}       & $\circ$   & $\bullet$ & $\circ$   & $\times$  & $\times$  \\ \hline
		IEMF/IEMA  \cite{45IEMFIEMA}  & $\circ$   & $\bullet$ & $\circ$   & $\times$  & $\times$  \\ \hline
		ROAM       \cite{46ROAM}      & $\circ$   & $\circ$   & $\circ$   & $\bullet$ & $\bullet$ \\ \hline
		LMR        \cite{47LMR}       & $\times$  & $\bullet$ & $\circ$   & $\bullet$ & $\bullet$ \\ \hline
		GRAB       \cite{48GRAB}      & $\circ$   & $\bullet$ & $\circ$   & $\times$  & $\bullet$ \\ \hline
		HMRP       \cite{49HMRP}      & $\bullet$ & $\circ$   & $\circ$   & $\times$  & $\bullet$ \\ \hline
		CBMPR      \cite{50CBMPR}     & $\circ$   & $\circ$   & $\circ$   & $\bullet$ & $\bullet$ \\ \hline
		DGR        \cite{51DGR}       & $\bullet$ & $\bullet$ & $\circ$   & $\times$  & $\bullet$ \\ \hline
		DCF        \cite{52DCF}       & $\bullet$ & $\bullet$ & $\circ$   & $\times$  & $\bullet$ \\ \hline
		RPL        \cite{53RPL}       & $\bullet$ & $\bullet$ & $\circ$   & $\times$  & $\bullet$ \\ \hline
		SAR        \cite{54SAR}       & $\circ$   & $\times$  & $\circ$   & $\times$  & $\times$  \\ \hline
		SPEED      \cite{55SPEED}     & $\circ$   & $\times$  & $\circ$   & $\bullet$ & $\times$  \\ \hline
		MMSPEED    \cite{56MMSPEED}   & $\circ$   & $\times$  & $\circ$   & $\bullet$ & $\bullet$ \\ \hline
		MGR        \cite{57MGR}       & $\bullet$ & $\circ$   & $\circ$   & $\times$  & $\times$  \\ \hline
		TERP       \cite{58TERP}      & $\circ$   & $\bullet$ & $\bullet$ & $\bullet$ & $\times$  \\ \hline
		SEECH      \cite{59SEECH}     & $\circ$   & $\circ$   & $\circ$   & $\times$  & $\times$  \\ \hline
		M-DART     \cite{60MDART}     & $\circ$   & $\bullet$ & $\bullet$ & $\bullet$ & $\bullet$ \\ \hline
		SPastry    \cite{61SPastry}   & $\circ$   & $\times$  & $\bullet$ & $\bullet$ & $\times$  \\ \hline
		VCP        \cite{62VCP}       & $\circ$   & $\times$  & $\bullet$ & $\bullet$ & $\times$  \\ \hline
	\end{tabular}
\end{table}

The category of location-based protocols consists of several algorithms that are possibly suitable for use in PCNs.
But since payment channels are location-independent, there is not an intuitive way to map nodes on specific geographic positions.
Therefore, the algorithms GEAR~\cite{35GEAR}, GEM~\cite{36GEM}, IGF~\cite{37IGF}, SELAR~\cite{38SELAR}, GDSTR~\cite{39GDSTR}, OGF~\cite{41OGF}, PAGER-M~\cite{42PAGERM} and HGR~\cite{43HGR} are also not further regarded.
\clearpage
The following routing protocols remain:

\begin{description}
	\item[E-TORA~\cite{07ETORA}]
	The successor protocol of TORA~\cite{03TORA} uses a directed acyclic graph (DAG) to find possible routes between a sender and a receiver.
	During route creation and maintenance, it takes energy parameters into account, in order to not always use the same path and to prolong the network lifetime.
	
	\item[ZRP~\cite{08ZRP}]
	The Zone Routing Protocol (ZRP) is a hybrid approach using proactive and reactive routing.	
	With a radius of $\rho$ hops, the zone of neighboring nodes is defined.
	These are kept in a routing table.
	Is a recipient outside of the zone, then the reactive component uses route request and reply messages to discover the node.
	
	\item[ROAM~\cite{46ROAM}]
	ROAM is an on-demand routing algorithm that uses directed acyclic multi-paths.
	In contrast to other on-demand protocols, ROAM maintains routing tables for destinations and link costs, to prevent sending unnecessary search packets.
	Destinations that are not present in the table, are located using diffusion search.
	
	\item[TERP~\cite{58TERP}]
	This protocol is a successor of the AODV routing protocol that uses route request and reply messages to discover a path on-demand.
	For the selection of the route, it takes the energy level as well as the trust of nodes into account.
	The trust estimation is done by monitoring the operation of neighbor nodes.
	
	\item[CBMPR~\cite{50CBMPR}]
	The cluster-based multi-path routing (CBMPR) protocol uses a hierarchical approach with cluster heads that know all members and some neighbor cluster heads.
	By choosing multiple paths through different clusters, resilience and scalability of the network are raised.
	
	\item[M-DART~\cite{60MDART}]
	This proactive protocol uses a DHT-based approach to route packets on multiple possible paths.
	To react on changes in the network, a dynamic addressing layer is used on top of the existing communication layer.
\end{description}
Finally, we select the protocols E-TORA~\cite{03TORA}, TERP~\cite{58TERP} and M-DART~\cite{60MDART} to be implemented and evaluated in our work.
The reason for choosing E-TORA is its promising feature to use different paths on subsequent messages to extend the network lifetime.
This could be turned into the selection of paths that offer higher balances and lower fees in PCNs.
As the only algorithm that takes primarily trust into account, TERP is an interesting approach based on AODV that has already been proven to be suitable for use in a PCN by Hoenisch et al.~\cite{Hoenisch2018AODV}.
To round up the selection, the DHT-based protocol M-DART contributes proactive routing to the former on-demand approaches.

\section{Full Evaluation Results}
\begin{figure}[bt]
	\centering
	\captionsetup{width=\linewidth}
	\includegraphics[width=\textwidth]{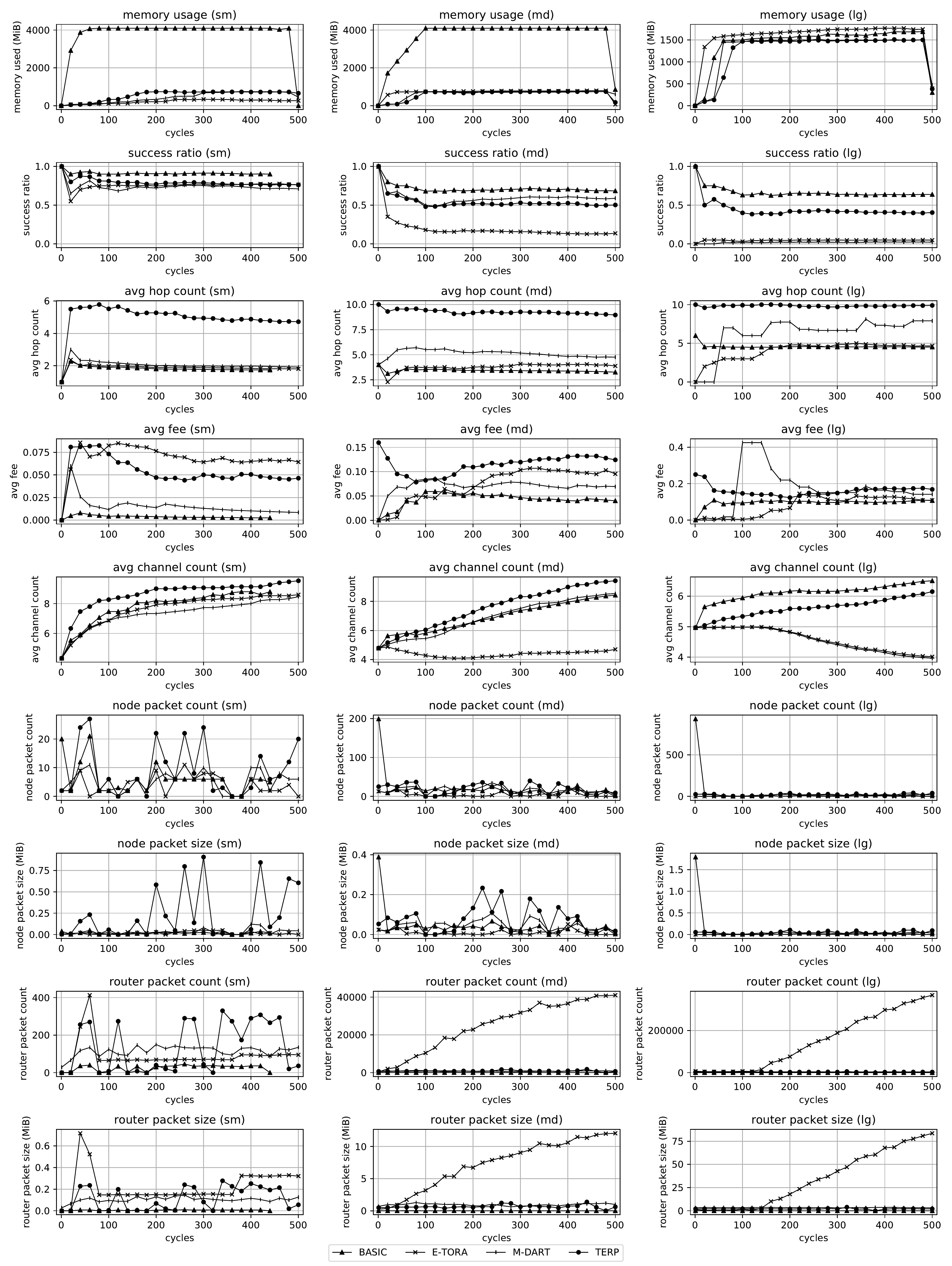}
	\caption{Complete Results of the Basic Scenario}
	\label{fig:app:raw:basic}
\end{figure}

\begin{figure}[bt]
	\centering
	\captionsetup{width=\linewidth}
	\includegraphics[width=\textwidth]{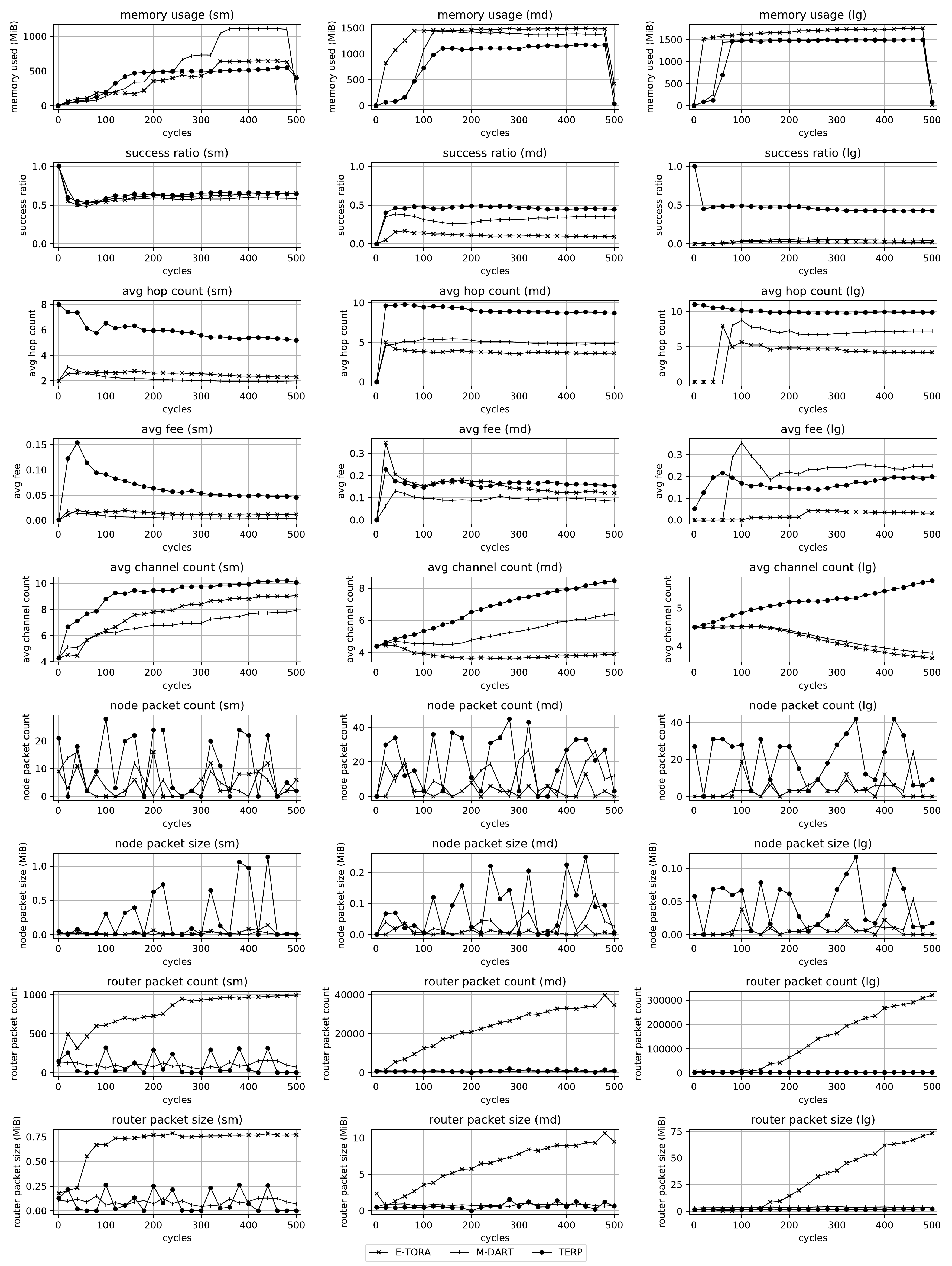}
	\caption{Complete Results of the Faulty Scenario}
	\label{fig:app:raw:faulty}
\end{figure}

\begin{figure}[bt]
	\centering
	\captionsetup{width=\linewidth}
	\includegraphics[width=\textwidth]{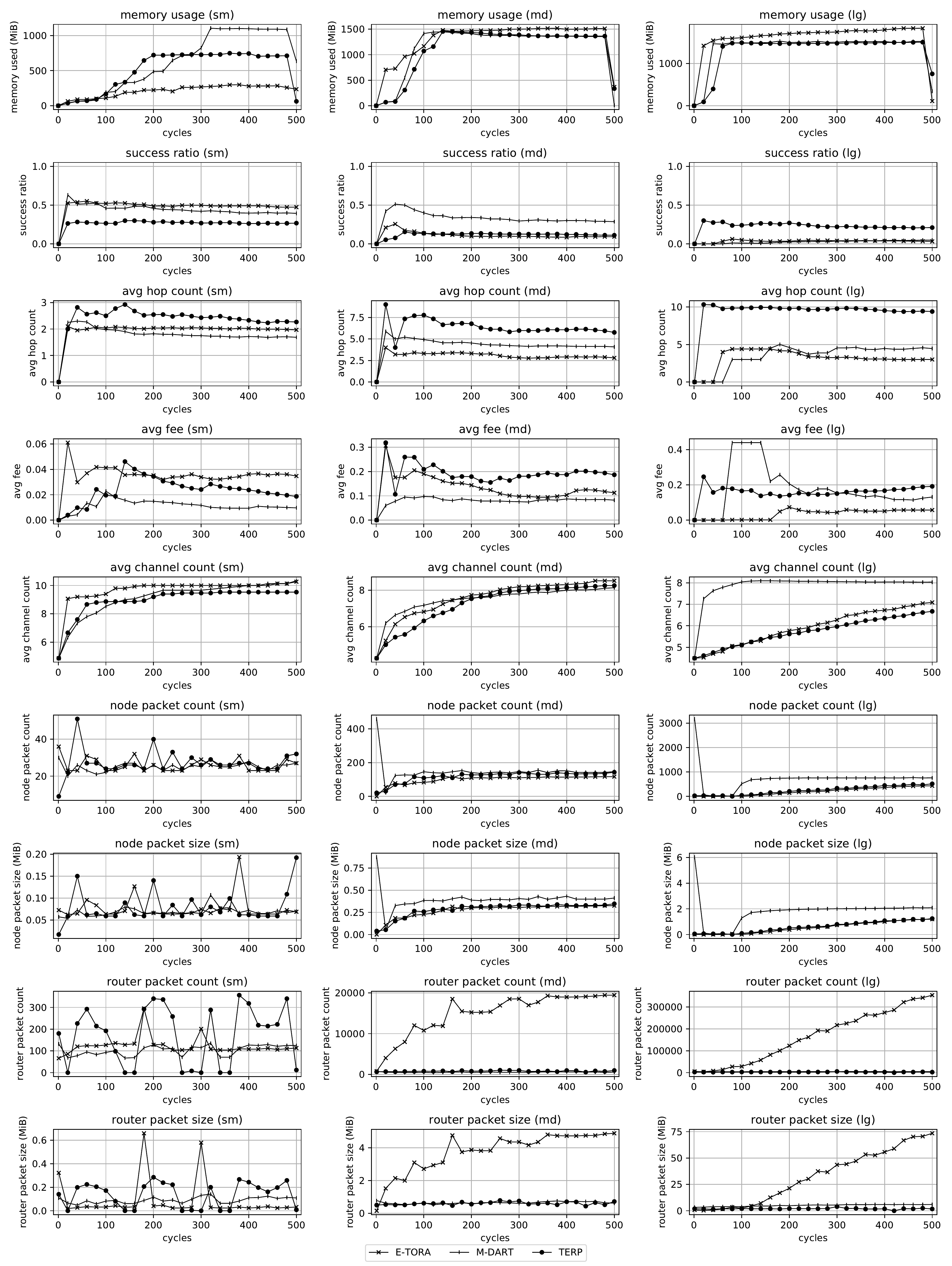}
	\caption{Complete Results of the Malicious Scenario}
	\label{fig:app:raw:malicious}
\end{figure}

\begin{figure}[bt]
	\centering
	\captionsetup{width=\linewidth}
	\includegraphics[width=\textwidth]{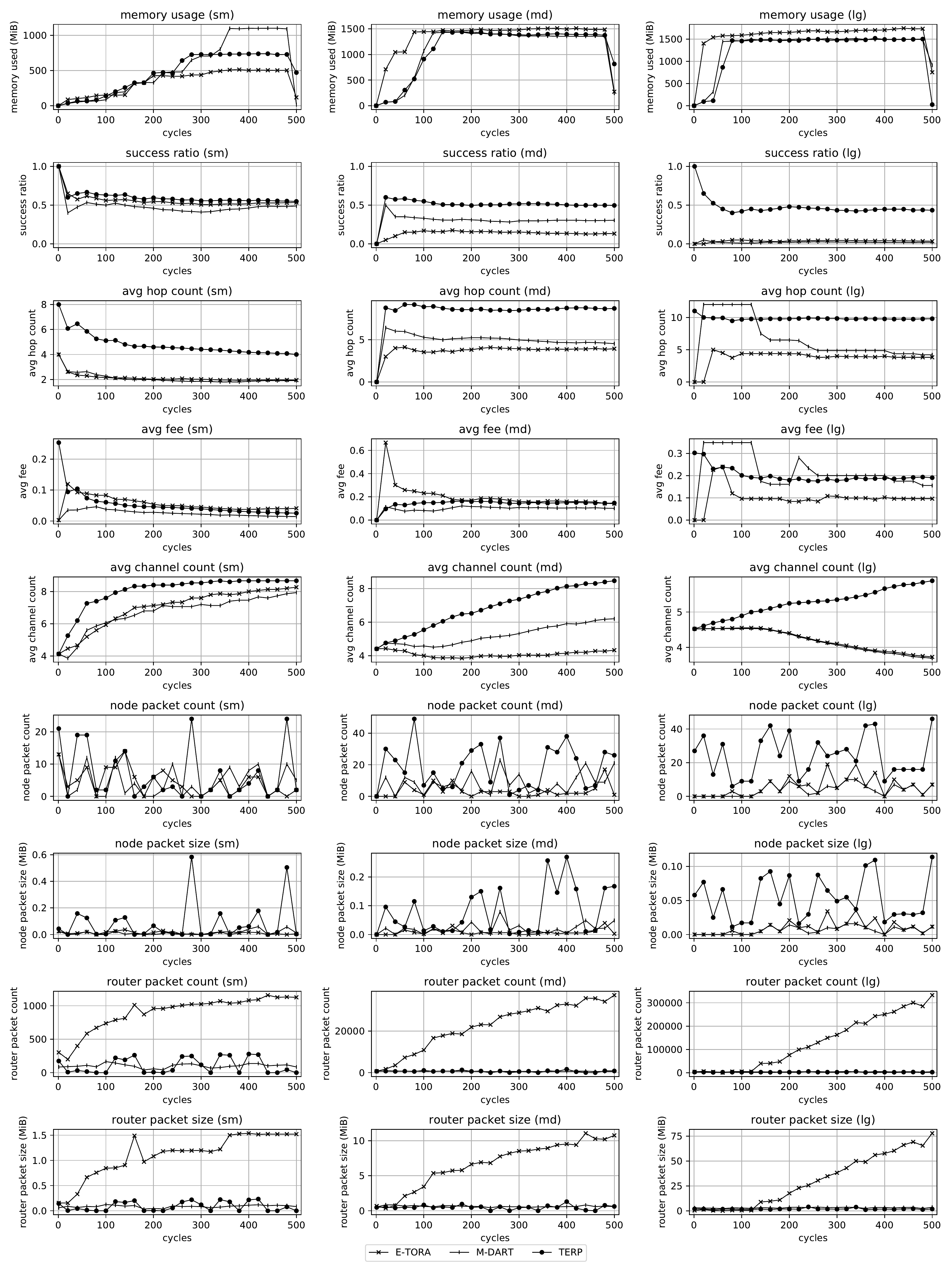}
	\caption{Complete Results of the Low-Participation Scenario}
	\label{fig:app:raw:lowpart}
\end{figure}

\begin{figure}[bt]
	\centering
	\captionsetup{width=\linewidth}
	\includegraphics[width=\textwidth]{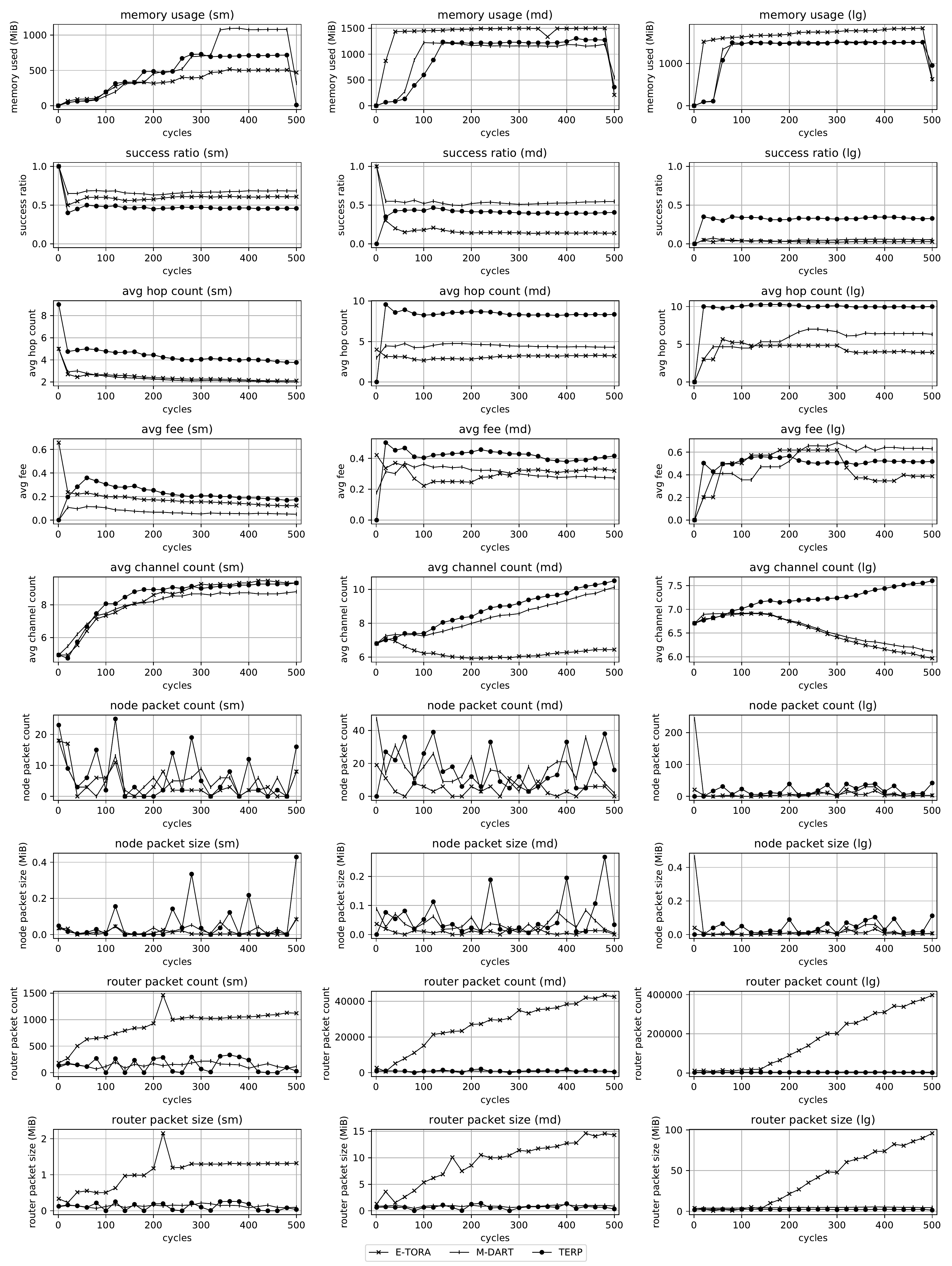}
	\caption{Complete Results of the Hub Scenario}
	\label{fig:app:raw:hub}
\end{figure}

\begin{figure}[bt]
	\centering
	\captionsetup{width=\linewidth}
	\includegraphics[width=\textwidth]{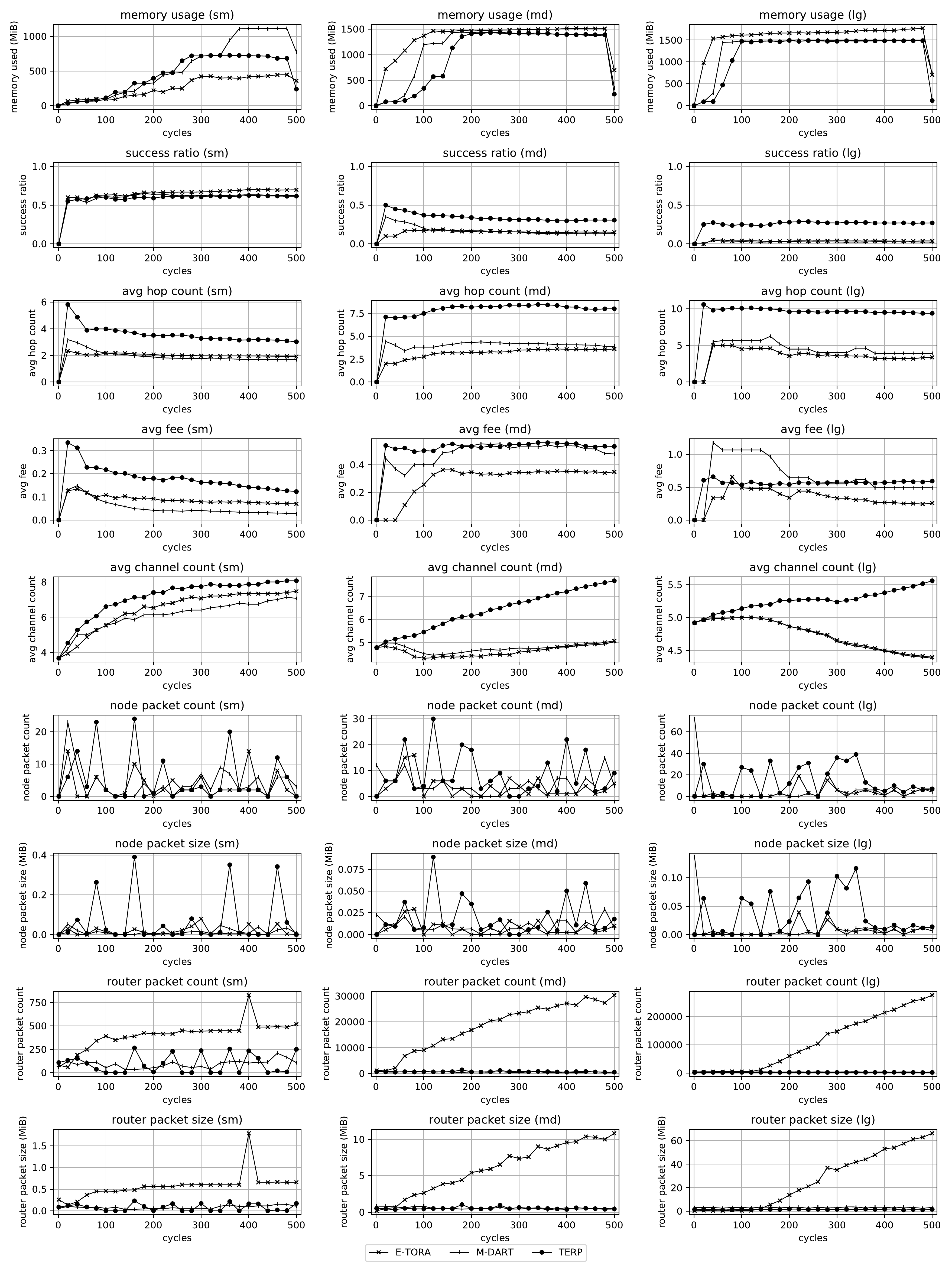}
	\caption{Complete Results of the Commercial Scenario}
	\label{fig:app:raw:commercial}
\end{figure}

In the following, we provide the full evaluation results with regard to memory consumption, success ratio, average hop count, average fee, average channel count, node packet count, node packet size, router packet count, and router packet size for the three implemented algorithms E-TORA, TERP, and M-DART. Results are shown for the basic scenario (Figure~\ref{fig:app:raw:basic})\footnote{Please note: For the purposes of preliminary evaluation runs, for the basic scenario, we have developed a BASIC routing protocol, which is based on unrealistic assumptions, e.g., a shared data structure to exchange network information and to always have an overview about nodes and channels in all parts of the network. Since the results are unrealistically good, we do not show them for all other scenarios.}, faulty scenario (Figure~\ref{fig:app:raw:faulty}), malicious scenario (Figure~\ref{fig:app:raw:malicious}), low-partipation scenario (Figure~\ref{fig:app:raw:lowpart}), hub scenario (Figure~\ref{fig:app:raw:hub}), and commercial scenario (Figure~\ref{fig:app:raw:commercial}). Also, different network sizes are taken into account: small (sm) with 30 nodes, medium (md) with 200 nodes, and large (lg) with 1000 nodes.

\clearpage
\begin{acks}
The financial support by the Austrian Federal Ministry for Digital and Economic Affairs, the National Foundation for Research, Technology and Development as well as the \grantsponsor{1}{Christian Doppler Research Association}{https://www.cdg.ac.at/} for the \grantnum{1}{Christian Doppler Laboratory for Blockchain Technologies for the Internet of Things} is gratefully acknowledged.
\end{acks}

\bibliographystyle{ACM-Reference-Format}
\bibliography{refs}

\end{document}